# A Portable In-Ear Pulse Wave Measurement System

R. Kusche[1], A. Malhotra[1], M. Ryschka[1], S. Kaufmann[1,2],

[1] Laboratory for Medical Electronics, Lübeck University of Applied Sciences, Germany, kaufmann@fh-luebeck.de
[2] Graduate School for Computing in Medicine and Life Sciences, University of Lübeck, Germany

## Abstract

The measurement of the pulse wave has proven to be a vital tool in medical diagnosis. Whereby most pulse wave measurements are carried out at extremities, this work proposes a system for measuring the pulse wave and the Pulse Arrival Time (PAT) in the interior of the ear. The developed measurement device is based on a battery powered microcontroller system. The measurement device simultaneously acquires an Einthoven Lead II Electrocardiogram (ECG), a dual wavelength Photoplethysmogram (PPG), the pressure in both ears, the temperature inside the ear canal, as well as the subject's motion. The acquired measurement data can either be saved on a micro SD-card or can be transmitted wireless via Bluetooth or wired via USB to a host PC for further analysis. Battery powered the device can operate up to 8 hours. In addition to the system description, first measurements carried out with the system will be presented.

## 1   Introduction

A key to future personal monitoring is portability and adaptability. For that purpose it is desirable to record vital signs non-invasively and comfortably for a long period of time.

Pulse Arrival Time (PAT) and the morphology of the pulse wave are considered as indicators of arterial stiffness and are also known as (prognostic) markers for cardiac and vascular diseases [1, 2, 3, 4, 5]. In the past various methods and sites have been demonstrated for pulse wave measurements, mostly on carotid and femoral artery through invasive and non-invasive methods.

This work proposes a system for detecting the pulse wave inside the auditory canal for PAT and Pulse Wave Velocity (PWV) measurements, as proposed by [6]. The developed measurement device is based on a battery powered microcontroller system and can simultaneously acquire a single channel Einthoven Lead II Electrocardiogram (ECG), a dual wavelength Photoplethysmogram (PPG), the pressure in both the ears, the temperature inside the ear canal, as well as the subject's motion via an accelerometer. After acquisition, the measured data can be transmitted via USB or wirelessly via Bluetooth to a host PC for further analysis. Additionally it is also possible to save the measurement data on a micro SD-card.

## 2   Measurement Methods

### 2.1   Electrocardiogram (ECG)

Via the recording of the Electrocardiogram (ECG) it is possible to detect the electrical activity of the heart, which is closely correlated with the contractions of the heart's atria and ventricles and can therefore provide a reliable time reference for the PPG and pressure measurements.

### 2.2   Photoplethysmography (PPG), Pulse Arrival Time (PAT) and Pulse Wave Velocity (PWV)

Caused by the rhythmic contraction of the heart, blood is pumped into the arterial system where it propagates as a pulse wave due to the elasticity of the arteries. The blood pulsation is time and site dependent and can be measured in terms of flow, pressure and volume. PPG simply detect the changes in volume of blood below the sensor by exploiting the absorption of light. The PPG in this work is recorded from the A. temporalis superficialis on the height of the Tragus region.

The morphology of the obtained waveform, as well as thereof derived PAT are very significant and direct prognostic markers for the stiffness of the arteries [1, 2, 3, 4, 5]. Hence provides vital information regarding the overall condition of the cardiovascular system. PAT values can also be used for the calculation of the PWV, which is also used as an important marker for arterial stiffness [2, 3].

The PAT is defined as the time span between the R-peak of the ECG and the arrival of the pulse wave at a specific position. The PAT is commonly defined as one of three possibilities according to Figure 1.

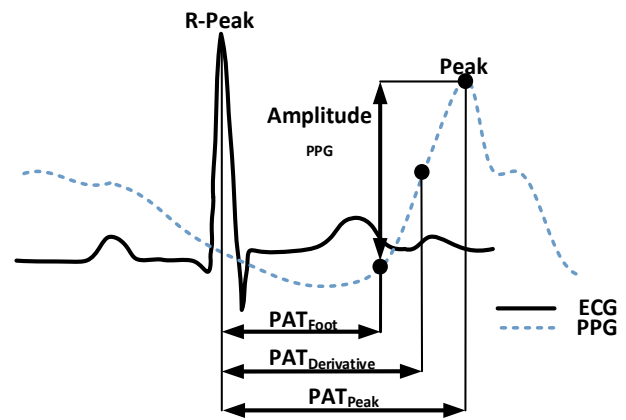

**Figure 1.** Calculation principle for the Pulse Arrival Time (PAT).

The PWV (1) is given by the ratio between the distance of two measurement sites (Δx) and the Pulse Transient Time (PTT) [7], which is the time required by the pulse wave to travel that distance.

$$PWV = \frac{\Delta x}{PTT} \qquad (1)$$

The PTT can also be expressed in terms of the PAT and the Pre-Ejection Period (PEP) according to (2). Where the PEP is the time between the R-peak of the ECG and the opening of the aortic valve.

$$PTT = PAT - PEP \qquad (2)$$

## 2.3 In-Ear Pressure

For the measurement of the internal pressure variation the auditory canal has to be sealed against the ambient pressure. The sealing in this work is achieved via stethoscope ear olives connected to a pressure sensor. Inside the sealed cavity a pressure variation can be observed [6]. According to the ideal gas law this pressure variation corresponds to a volume change, under the assumption of a constant temperature. Figure 2 shows the principle drawing of the in-ear pressure model of the auditory canal.

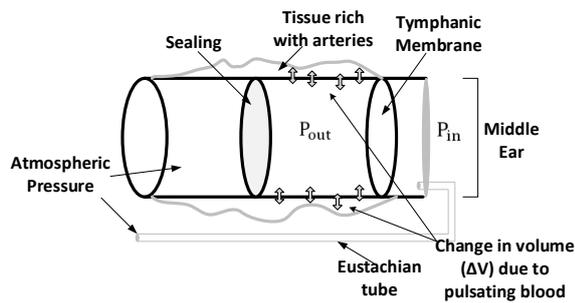

**Figure 2**. Basic model of the pressure changes in the auditory canal.

The volume change can in principle be caused by two effects or their combination 1) the blood vessels around the auditory canal expand during the inrush of the pulse wave 2) the tympanic membrane moves corresponding to the heart beat.

## 2.4 Acceleration and Temperature

For the assessment of the subject's position and physical activity an accelerometer is implemented in the measurement system. Additionally the temperature in the ear canal can be measured via a NTC resistor based temperature sensor.

## 3 Implementation

### 3.1 Block diagram

Figure 3 shows a block diagram of the developed portable in-ear pulse wave measurement system.

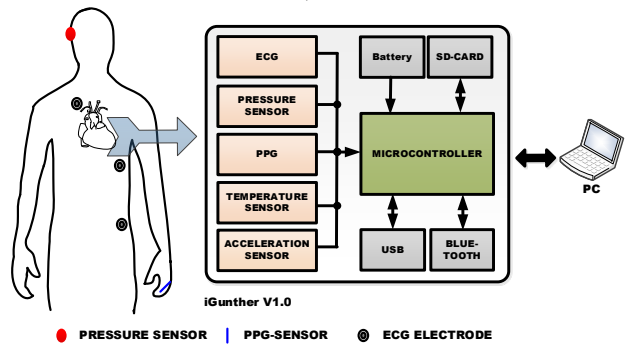

**Figure 3.** Block diagram of the portable microcontroller based measurement system.

The microcontroller based embedded system acquires all measurement data from the subject and transmit the measured data to the host PC for further analysis and display. The data can be transmitted via Bluetooth, USB or can be stored on the micro SD-card.

### 3.2 ECG Module

The ECG module consists of an Instrumentation Amplifier (INA) with a Driven Right Leg (DRL), a driven shield, and a base line wandering rejection circuit. The ECG module measures in a frequency range of 150 mHz to 150 Hz. The analogue ECG signal is digitized via the internal 12 bit Analogue to Digital Converter (ADC) of the microcontroller.

### 3.3 Photoplethysmography (PPG) Module

The implemented PPG circuit is designed for reflective mode and is based on two time multiplexed LEDs with wavelengths of 660 nm and 940 nm respectively. The reflected light is sensed by a photodiode, which is connected to a transimpedance amplifier (OPA2380 from Texas Instruments). Whose output signal is used on one hand to control the current of the LEDs by a digital control loop to ensure the reflected light intensity is in the optimal reception range of the detector circuit. On the other hand the signal is band-pass filtered with a pass-band of ≈ 15 mHz to ≈ 30 Hz to remove the DC-offset due to the dark current of the photo diode and to mitigate 100 Hz flickering of the ambient light. The cut-off frequencies are also chosen to ensure the fidelity of the desired PPG waveform, whose spectral components are ranged mainly from a few mHz up to ≈ 20 Hz. Finally the PPG signal is digitized via another channel of the microcontroller's ADC.

### 3.4 Pressure Measurement

The pressure sensors (HCE-M010DBE8P3 from First Sensor AG) have a calibrated and compensated pressure measurement range of ±1000 Pa with a digital Serial Peripheral Interface (SPI) output. The pressure sensor supports up to 1 kSPS with 14 bit precision, which leads to a theoretic resolution of 122 mPa.

### 3.5 Acceleration Measurement

The acceleration measurement is based on a LIS3DH (from ST Microelectronics) accelerometer. The sensor can measure in three directions with full scale ranges from ±2 g up to ±16 g with 16 bit precision per direction at a sample rate from 1 SPS to 5 kSPS and is connected to the microcontroller via the SPI interface.

### 3.6 Temperature Measurement

The temperature measurement is based on a NTC resistor sensor with dimensions of 1.0 x 0.5 x 0.5 mm$^3$, which allows the sensor be located inside the ear canal. The NTC resistance is evaluated with a quarter Wheatstone bridge and its calibration is optimized for the temperature range around 37 °C. The output signal of the bridge is amplified and digitized with a further channel of the internal ADC of the microcontroller.

### 3.7 Housing and Battery

The housing for the measurement system was designed in SolidWorks (Dassault Systems) and manufactured with a 3d-printer (MakerBot Replicator 2X). The housing consists of ABS plastic and has a size of 71.5 x 71.5 x 38 mm$^3$. Figure 4 shows an exploded assembly drawing of the housing inclusive the Printed Circuit Board (PCB) and the used battery.

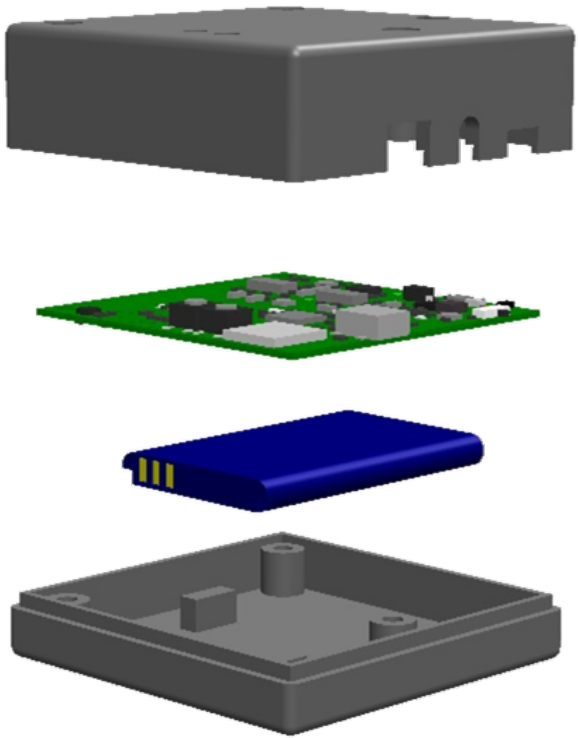

**Figure 4.** Exploded assembly drawing of the housing of the measurement system inclusive the Printed Circuit Board (PCB) and the used battery.

The li-ion battery has 3.7 V by 1.25 Ah with dimensions of about 53 x 34 x 5.5 mm$^3$ and weighs about 23 g. The expected run-time with the battery is about 8 hours.

### 3.8 Microcontroller System and Communication Interfaces

The microcontroller (ATxMega128A4U from Atmel Corp.) acquires the measurement data from the different sensors and handles data transmission to the host PC via Bluetooth or USB. Optionally the measurement data can be stored on the micro SD-card. While USB communication is realized via the internal USB stack of the microcontroller, Bluetooth is implemented with a commercially available module (RN42-I/RM from Roving Networks). The Bluetooth module is certified according to Bluetooth V2.1 incl. Enhanced Data Rate (EDR) mode and supports master and slave mode with up to 300 kbps. The interfacing to the microcontroller is implemented as asynchronous serial connection. Alternatively to the battery the system can be supplied externally via USB. Whereas the digital components operate with +3.3 V only, the analogue components are supplied with ±3.3 V. For electrical safety considerations it is recommended to operate the system without USB connection in wireless or SD-card storage mode while running from the system battery.

### 3.9 Manufactured System

Figure 5 shows a photograph of the manufactured and populated Printed Circuit Board (PCB) of the measurement system. The PCB has a size of about 60 x 60 mm$^2$ and contains about 200 components.

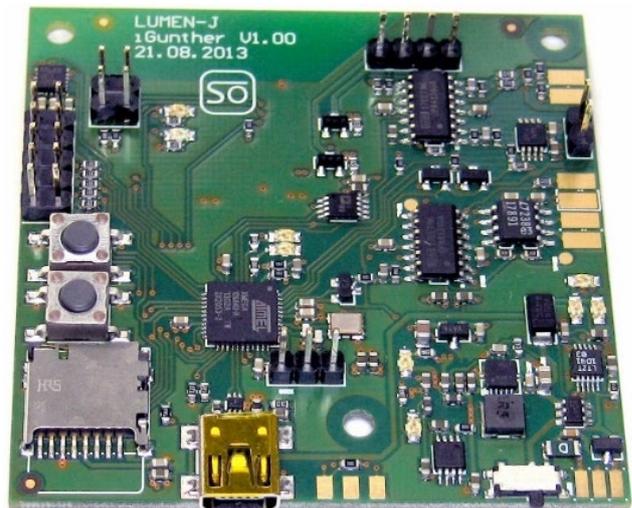

**Figure 5.** Manufactured and populated Printed Circuit Board (PCB) of the developed measurement system. The PCB has dimensions of 60 x 60 mm$^2$ and contains about 200 components.

### 3.10 Software

The firmware of the measurement system is based on the Atmel Software Framework (ASF) and is written in C language.

The PC interface software is written in C# language and is based on the .NET framework from Microsoft. It is able to display the measured waveforms in real-time with a latency of about 50 ms and can be used to configure the embedded measurement systems in terms of active channels and sample rates. With the interface software, waveforms can be recorded and exported to MathWorks MATLAB or Microsoft Excel. Figure 6 shows an image of the Graphical User Interface (GUI) of the software with the acquired waveforms. Based on the recorded PPG, ECG and pressure waveforms, PAT and PWV values are derived.

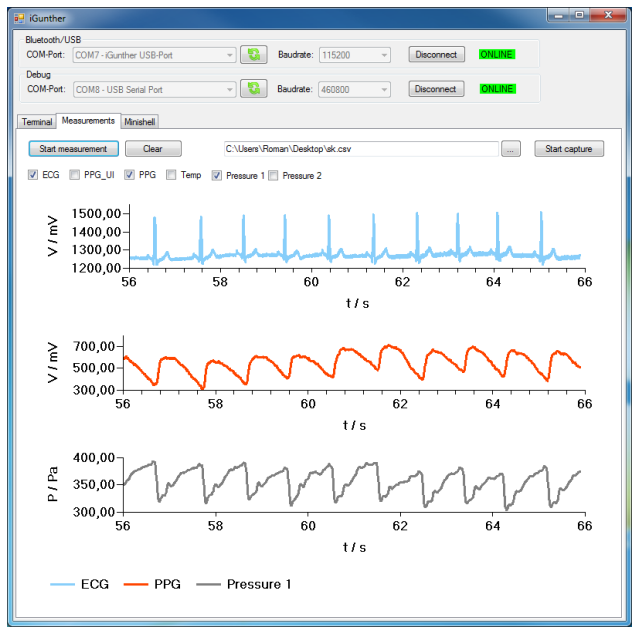

**Figure 6.** Screenshot of the user interface of the developed C# based PC software.

## 4  Results and Discussion

Table 1 shows preliminary measurement results acquired over 30 heart beats on two healthy male subjects. For simplicity reasons only measurements on the left ear are shown. The measurement signals are sampled with 1 kSPS and are digitally processed with zero-phase filters in MathWorks MATLAB. Afterwards peak detections and derivation calculations are performed on the filtered signals. The PPG and pressure waveforms are referenced to the ECG's R-peak for evaluating the PAT values. The PAT derived from the pressure measurements are in a range of about 150 ms to 170 ms and the PAT derived from the PPG measurements are in a range of about 190 ms to 210 ms. These results are reasonable taking into account that usually reported PAT values on extremities are in the range of 240 ms [8]. With an assumed PEP of 70 ms and an estimated time difference of 50 ms between $PAT_{Foot}$ and $PAT_{Derivative}$ the results of the pulse wave velocities are $PWV_{PPG} \approx 3.8$ m/s and $PWV_{Pressure} \approx 7.4$ m/s for $\Delta x = 30$ cm. This indicates that the pressure wave in the ear measured through a pressure sensor occurs before the change of blood volume in the arteries, measured via PPG.

## 5  Conclusion

A prototype of an in-ear pulse wave measurement system was developed and tested. The results are very promising and showing a good performance. In future the prototype has to be miniaturized and the interface software must be enhanced by the possibilities to automatically measure PAT and PWV, as well as the heart rate. Furthermore a second PPG channel for measurements on the extremities should be added to have an additional reference measurement side for comparisons.

|  | Subject 1 | Subject 2 |
|---|---|---|
| **Age** | 27 | 29 |
| **Sex** | male | male |
| **Height** | 190 cm | 188 cm |
| **Weight** | 92 kg | 134 kg |
| **Heart Rate** | 86 BPM | 64 BPM |
| **Mean** $PAT_{Derivative, PPG}$ | 197 ms | 201 ms |
| **Mean** $PAT_{Derivative, Press.}$ | 154 ms | 167 ms |

**Table 1.** Results of the PAT measurements

## 6  Acknowledgment

The authors would like to thank F. Adam, G. Ardelt, K. Breßlein, and N. Hunsche for supporting this work and Linear Technology and Texas Instruments for the provision of free samples during the development process.
This publication is a result of the ongoing research within the LUMEN research group, which is funded by the German Federal Ministry of Education and Research (BMBF, FKZ 13EZ1140A/B).